\documentclass{ifacconf}

\usepackage{natbib}        %

\usepackage[utf8]{inputenc}
\usepackage{courier}
\usepackage{cite}
\usepackage{graphicx}
\usepackage[cmex10]{amsmath}
\usepackage{array}
\usepackage{color}
\usepackage{amssymb}
\usepackage{amsfonts}
\usepackage[short]{optidef}
\usepackage{balance}
\usepackage{subcaption}
\usepackage{float}
\usepackage{subfiles}
\usepackage{svg}
\usepackage{comment}
\usepackage{multirow}
\usepackage{multicol}
\usepackage{arydshln}
\def\BibTeX{{\rm B\kern-.05em{\sc i\kern-.025em b}\kern-.08em
    T\kern-.1667em\lower.7ex\hbox{E}\kern-.125emX}}

\usepackage{physics}

\newcommand{\uk}{u_k}
\newcommand{\xbase}{x_\mathrm{b}}
\newcommand{\xbasek}{x_{\mathrm{b},k}}
\newcommand{\xbasekplus}{x_{\mathrm{b},k+1}}

\newcommand{\yestk}{\hat y_k}

\begin{document}

\begin{frontmatter}

\title{Encoder initialisation methods in the model augmentation setting\thanksref{footnoteinfo}} 

\thanks[footnoteinfo]{This work is funded by the European Union (Horizon Europe, ERC, COMPLETE, 101075836) and by the Air Force Office of Scientific Research under award number FA8655-23-1-7061. Views and opinions expressed are however those of the author(s) only and do not necessarily reflect those of the European Union or the European Research Council Executive Agency. Neither the European Union nor the granting authority can be held responsible for them.}

\author[First]{J.H. Hoekstra} 
\author[Second]{B. Gy\"or\"ok}
\author[First,Second]{R. T\'{o}th}
\author[First]{M. Schoukens}

\address[First]{Control Systems Group, Dept. of Electrical Eng., Eindhoven University of Technology, Eindhoven, The Netherlands.}
\address[Second]{Systems and Control Laboratory, HUN-REN Institute for Computer Science and Control, Budapest, Hungary}

\begin{abstract}
    \emph{Nonlinear system identification}~(NL-SI) has proven to be effective in obtaining accurate models for highly complex systems. Recent encoder-based methods for \emph{artificial neural network state-space}~(ANN-SS) models have shown state-of-the-art performance with improved computational efficiency, where the encoder is used to estimate the initial state allowing for batch optimisation methods. To address the lack of interpretability of these black-box ANN models, model augmentation approaches can be used. These combine prior available baseline models with the ANN learning components, resulting in faster convergence and more interpretable models. The combination of the encoder-based method with model augmentation has shown potential. Thus far, however, the encoder has still been treated as a black-box function in the overall estimation process, while additional information in the form of the baseline model is available to predict the model state from past input-output data. In this paper, we propose novel encoder initialisation approaches based on the available baseline model, resulting in improved noise robustness and faster convergence compared to black-box initialisation. The performance of these initialisation methods is demonstrated on a mass-spring-damper system.

\end{abstract}

 \label{sec:Abstract}

\end{frontmatter}

\section{Introduction} \label{sec:Introduction}
As control systems are becoming more complex and performance requirements surge, the need for accurate nonlinear models capable of efficiently capturing complicated behaviours of physical systems is rapidly increasing. It is common practice to derive physics-based models using \emph{first-principle} (FP) methods, e.g., rigid-body dynamics \citep{spong2006robot}; however, these models provide only an approximate system description. Although more accurate FP models can be developed, this is a labour-intensive process, especially when additional physical effects are included, such as friction or aerodynamic forces. Modelling these phenomena from first principles often requires dedicated experimental campaigns to identify and estimate the associated unknown parameters. Furthermore, the resulting models may become too complex to handle analytically. In some cases, reliable FP descriptions for such effects may not even exist, limiting the modelling process to approximations with varying levels of fidelity.

To overcome these issues, \emph{nonlinear system identification} (NL-SI) methods offer an alternative option to estimate models directly from measurement data \citep{Schoukens2019}. Black-box models, particularly those that incorporate \emph{artificial neural networks} (ANNs), have achieved unprecedented accuracy in capturing complex behaviours. In control applications, ANN-based \emph{state-space} (SS) models have proven to be effective in handling high-order systems and capturing complex nonlinear dynamics \citep{suykens1995, Schoukens2021}. Specifically, encoder-based methods \citep{masti2021learning, beintema2024data} have made the identification problem computationally feasible. These methods rely on splitting the dataset into smaller segments, allowing for efficient batch optimisation methods. It is common practice to treat the initial states as optimisation variables, which in the splitting of the dataset would result in a large increase in the number of parameters. The encoder overcomes this problem by estimating these initial states from past input-output data.

While these methods are accurate and computationally efficient, they do result in black-box models. These are difficult to interpret, which limits the reliability of the model during, e.g., extrapolation beyond the training data.
This is a significant drawback for control applications, where interpretable models are preferred in the design process \citep{ljung2010perspectives}. The second drawback is the significant time spent learning expected behaviour that has already been modelled thoroughly, e.g., FP-based understanding of the rigid-body dynamics of the system. This can be addressed through model augmentation methods \citep{sun2020comprehensive, gotte2022composed, Groote2022, Shah2022}. This method combines physics-based, e.g., baseline, models with flexible function approximators, such as ANNs, in a combined model structure. As a result of this structural combination, the already known knowledge is captured in the baseline model, and the learning components only need to model the unknown dynamics. For control engineering, this approach further enhances the interpretability of the obtained model, as it is clear how the well-understood baseline model is augmented and how this affects the behaviour of the model.

The combination of model augmentation with encoder-based methods has been shown to result in accurate models with fast convergence \citep{Retzler2024, HOEKSTRA2025, kessels2025ai}. However, these method treat the encoder as a randomly initialised ANN \citep{Retzler2024, HOEKSTRA2025} or assume the states measured to some degree \citep{kessels2025ai}. There is, however, additional prior information available in the form of a baseline model, which could be utilised to initialise the encoder. Previous work has shown that initialising the encoder and the state progression network based on a \textit{best linear approximation} (BLA) can improve the convergence time in the ANN-SS identification setting \citep{ramkannan2023initialization}. To the best of the authors' knowledge, no such encoder initialisation method has been proposed for a nonlinear baseline model.

This paper proposes such encoder initialisation methods based on the prior (nonlinear) baseline models. First, an LTI model-based method is proposed, which is expanded to include noise models. This LTI method is based on state reconstruction theory directly. Such model-based methods work well for LTI models. For nonlinear models, reconstruction theory relies on intractable function inversions. This paper proposes, first, a linearisation of the nonlinear model to address this. As an alternative approach, we propose a data-based method, estimating the ANN encoder prior to the SS-ANN estimation based on the baseline model.

The paper is organised as follows: first, the system identification problem and model augmentation approach are presented in Section~\ref{sec:Problem}. Then, Section~\ref{sec:Method} introduces the model-based and data-based encoder initialisation methods. A hardening \emph{mass-spring-damper} (MSD) simulation example is used to analyse and compare the performance of the proposed encoder initialisation methods in Section~\ref{sec:Example}. The conclusions are given in Section~\ref{sec:Conclusion}.

\section{Identification Problem} \label{sec:Problem}
As the true system to be identified, we consider the data generating system given by the \emph{discrete-time} (DT) nonlinear representation
\begin{subequations}\label{eq:nl_dyn}
	\begin{align}
		x_{k+1}&=f(x_k,u_k, e_k),\\
		y_k&=h(x_k, u_k) +e_k,
	\end{align} 
\end{subequations} 
where $x_k \in \mathbb{R}^{n_\mathrm{x}}$ is the state, $u_k \in \mathbb{R}^{n_\mathrm{u}}$ is the input, $y_k \in \mathbb{R}^{n_\mathrm{y}}$ is the output signal of the system at time moment $k\in\mathbb{Z}$ with $e_k$ an i.i.d. white noise process with finite variance, $f: \mathbb{R}^{n_\mathrm{x}} \times \mathbb{R}^{n_\mathrm{u}} \rightarrow \mathbb{R}^{n_\mathrm{x}}$ is the state-transition function and $h: \mathbb{R}^{n_\mathrm{x}}\rightarrow \mathbb{R}^{n_\mathrm{y}}$ is the output function. This state-space representation is a general form that can describe a wide range of system dynamics and noise scenarios encountered in practice.

We assume a stable baseline model of \eqref{eq:nl_dyn} is available in an NL-SS form
\vspace*{-4pt}
\begin{subequations}\label{eq:baseline_model}
    \begin{align}
        \xbasekplus & = f_\text{base}\left(\xbasek, \uk, e_k\right), \label{eq:baseline_state_transition}\\
        \yestk & = h_\text{base}\left(\xbasek, \uk\right),
    \end{align}
\end{subequations}
where $\xbasek \in \mathbb{R}^{n_{\xbase}}$ is the baseline model state, $\yestk \in \mathbb{R}^{n_\mathrm{y}}$ is the model output, %
and $f_\text{base} : \mathbb{R}^{n_{\xbase}} \times \mathbb{R}^{n_\mathrm{u}} \rightarrow \mathbb{R}^{n_{\xbase}}$ with $h_\text{base}: \mathbb{R}^{n_{\xbase}} \times \mathbb{R}^{n_\mathrm{u}}  \rightarrow \mathbb{R}^{n_{\mathrm{y}}}$ are the baseline state-transition and output readout functions respectively. The noise contribution in \eqref{eq:baseline_state_transition} can be removed if only a deterministic baseline model is available.

To model dynamic unmodeled terms, augmented states $x_{\mathrm{a},k}\in\mathbb{R}^{n_{x_\mathrm{a}}}$ are introduced in the following structure: $\hat{x}_k=[\xbasek^\top \ x_{\mathrm{a},k}^\top]^\top$. Now $\hat{x}_k$ denotes the combined model state of the augmentation structure. We consider a general model augmentation formulation \citep{HOEKSTRA2025}
\begin{subequations}\label{eq:augmentation_structure}
    \begin{align}\label{eq:interconn_mx}
        \hspace*{-2.9mm}  
        \left[\begin{array}{c}
        \hat x_{k+1} \\
        \hat y_k \\
        \hdashline[5pt/2pt]
        z_{1,k} \\
        z_{2,k} \\
        \end{array}\right] & \hspace*{-1mm} = \hspace*{-1mm}
        \underbrace{\left[\begin{array}{ccc;{5pt/2pt}cc}
        \mathit{S_{x x}} & \mathit{S_{x u}} & \mathit{S_{x e}} & \mathit{S_{x w_1}} & \mathit{S_{x w_2}}\\
        \mathit{S_{y x}} & \mathit{S_{y u}} & \mathit{S_{y e}} & \mathit{S_{y w_1}} & \mathit{S_{y w_2}}\\
        \hdashline[5pt/2pt]
        \mathit{S_{z_1 x}} & \mathit{S_{z_1 u}} & \mathit{S_{z_1 e}} & \mathit{S_{z_1 w_1}} & \mathit{S_{z_1 w_2}} \\
        \mathit{S_{z_2 x}} & \mathit{S_{z_2 u}} & \mathit{S_{z_2 e}} & \mathit{S_{z_2 w_1}} & \mathit{S_{z_2 w_2}} \\
        \end{array}\right]}_\mathbf{S}
        \hspace{-2pt} \hspace*{-1mm}  \left[\begin{array}{c}
            \hat x_k \\
            u_k \\
            e_k \\
            \hdashline[5pt/2pt]
            w_{1,k} \\
            w_{2,k} \\
        \end{array}\right] \\
        w_{1,k} & = \mathnormal{\phi}_\text{base}\left(z_{1,k}\right) = \begin{bmatrix}
            f_\text{base}\left(z_{1,k}\right)\\
            h_\text{base}\left(z_{1,k}\right)
        \end{bmatrix}\\
        w_{2,k} & = \mathnormal{\phi}_\text{aug}\left(z_{2,k}\right)\label{eq:phi_aug},
    \end{align}
\end{subequations}
where $\mathbf{S}\in \mathbb{R}^{n \times m}$ is the interconnection matrix, $\mathit{S}$ are selection matrices of appropriate size, and $\phi_\text{aug}$ are the learning components, parameterised by $\theta_\text{aug} \in\mathbb{R}^{n_{\theta_\text{aug}}}$, which is not shown in \eqref{eq:phi_aug} for notational clarity. %
The parametrisation of $\mathbf{S}$ can then represent a wide variety of model augmentation structures \citep{HOEKSTRA2025}. Similarly to the baseline model, the noise contribution can be removed from \eqref{eq:interconn_mx} to consider only the deterministic scenario.

We consider the identification algorithm from previous work by the authors \citep{HOEKSTRA2025} based on the work in \citep{beintema2024data}. The approach uses a multiple-shooting-based truncated objective function, where the input-output dataset $\mathcal{D}_N$, generated by \eqref{eq:nl_dyn}, is divided into $N$~segments of length~$T$, enables efficient batch optimisation and improved data efficiency:
\begin{subequations} \label{eq:loss_function}
    \begin{align}
        \hspace*{-0mm} V_\text{trunc}(\theta) \hspace*{-0mm} =  \frac{1}{2 N(T+1)} & \sum_{i=1}^N \sum_{\ell=0}^{T-1}\left\|\hat{y}_{k_i+\ell \vert k_i}-y_{k_i+\ell}\right\|_2^2 \label{eq:mse_loss}\\ 
    \begin{bmatrix}
        \hat{x}_{k_i+\ell + 1 \vert k_i} \\
        \hat{y}_{k_i+\ell \vert k_i}
    \end{bmatrix} &:=\mathnormal{\phi}_\theta\left(\theta_\text{aug}, \hat{x}_{k_i+\ell \vert k_i}, u_{k_i+\ell}, e_{k_i+\ell}\right) \label{eq:phi_theta}\\
    \hat{x}_{k \vert k} & :=\mathnormal{\psi}_{\theta}\left(\theta_\text{enc}, y_{k-n_a}^{k-1}, u_{k-n_b}^{k-1}\right),
    \end{align}
\end{subequations}
where $\theta=\mathrm{vec}(\theta_\mathrm{aug}, \theta_\mathrm{enc})$\footnote{It is possible to co-estimate the baseline model parameters with $\theta_\mathrm{aug}$ and $\theta_\mathrm{enc}$ in the model augmentation setting, see, e.g., \citep{HOEKSTRA2025}. This is, however, outside the scope of this paper.} is the joint parameter vector, $\phi_{\theta}$ is the LFR-based augmentation structure~\eqref{eq:augmentation_structure} constructed from $\phi_\text{base}$, $\phi_\text{aug}$ and $\mathbf{S}$, and $k+\ell \vert k$ indicates the state $\hat{x}_k$ or the output $\hat{y_k}$ at time $k + \ell$ simulated from the initial state $\hat{x}_{k \vert k}$ at time $k$. The subsections start at a randomly selected time $k_i\in \{n+1,\ldots,N-T\}$. The initial state of these subsections is estimated by an encoder function $\psi_\theta$ from past input-output data, i.e., $\hat{x}_{k|k}=\psi_\theta(y_{k-n_a}^{k-1}, u_{k-n_b}^{k-1}, e_{k-n_a}^{k-1})$ where $u_{k-n_b}^{k-1}= [\ u_{k-n_b}^\top \ \cdots \ u_{k-1}^\top ]^\top$ and $y_{k-n_a}^{k-1}$, $e_{k-n_a}^{k-1}$ are defined similarly. The noise realisation in \eqref{eq:phi_theta} is not directly available; it is typically estimated via the prediction error $\hat{e}_k=y_k-\hat{y}_k$. In the following, we consider the effect of process noise only on state reconstruction. The general formulation above could, for instance, be used to design an innovation noise filter during estimation, similar to \citep{beintema2024data}, which is beyond the scope of this work.

The existence of the encoder $\psi_\theta$ is proven by \citep{beintema2024data}. By rewriting the data generating system \eqref{eq:nl_dyn} as the n-step ahead predictor, we derive
\begin{subequations}
    \begin{align}
        y_k^{n+k} & = \Upsilon_n\left(x_k, u_k^{n+k}, e_k^{n+k}\right)\\
        & = \begin{bmatrix}
            h(x_k, u_k) + e_k \\
            (h \circ f)(x_k, u_k^{k+1}, e_k) + e_{k+1} \\
            \vdots \\
            (h \circ_n f)(x_k, u_k^{n+k}, e_k^{n+k-1}) + e_{n+k} \\
        \end{bmatrix}.
    \end{align}
\end{subequations}
where $\circ_n$ stands for $n$ times recursive function composition. If $\Upsilon_n$ is partially invertible as $x_k = \mathnormal{\Phi}_n(u_k^{n+k}, y_k^{n+k}, e_k^{n+k})$, then the reconstructability map \citep{isidori1985nonlinear} is given as
\begin{subequations}
    \begin{align}
        x_k & = (\circ_n f)(x_{k-n}, u_{k-n}^{k}, y_{k-n}^{k}) \\
        & = (\circ_n f)(\mathnormal{\Phi}_n(u_{k-n}^{k}, y_{k-n}^{k},e_{k-n}^{k}), u_{k-n}^{k}) \\
        & = \mathnormal{\Psi}_n(u_{k-n}^{k}, y_{k-n}^{k}, e_{k-n}^{k}).
    \end{align}
\end{subequations}
The noise sequence $e_{k-n}^{k}$ is not available in practice. Under the assumption that the noise $e_k$ is i.i.d. white noise, we can formulate the conditional expectation instead
\begin{equation}
    \bar x_k = \mathbb{E}_{e_k}\left[x_k | u_{k-n}^{k}, y_{k-n}^{k}\right] = \bar{\mathnormal{\Psi}}(u_{k-n}^{k}, y_{k-n}^{k}),
\end{equation}
which is an unbiased estimator of $x_k$. This estimator can be computed for LTI models, but is practically infeasible to compute for nonlinear models due to the required analytical inversion to obtain $\mathnormal{\Phi}_n$. Thus, the parameterised function estimator $\psi_\theta$ is used to approximate $\bar{\mathnormal{\Psi}}$. This encoder is co-estimated with $\mathbf{S}$, $\phi_\text{base}$ and $\phi_\text{aug}$.

We parameterise $\psi_\theta$ as a ResNet architecture in the form
\begin{multline}\label{eq:encoder_ann_struct}
    \begin{bmatrix}
        x_{\mathrm{b},k\vert k}\\x_{\mathrm{a},k\vert k}
    \end{bmatrix} = \psi_\theta\left(y_{k-n_a}^{k-1}, u_{k-n_b}^{k-1}\right) = \\\tilde{\psi}_{\tilde{\theta}}\left(y_{k-n_a}^{k-1}, u_{k-n_b}^{k-1}\right) + \begin{bmatrix}
        W_{\psi,y}^\mathrm{b} & W_{\psi,u}^\mathrm{b}\\
        W_{\psi,y}^\mathrm{a} & W_{\psi,u}^\mathrm{a}
    \end{bmatrix} \begin{bmatrix}
        y_{k-n_a}^{k-1}\\ u_{k-n_b}^{k-1}
    \end{bmatrix},
\end{multline}
where $\tilde{\psi}$ is a feedforward ANN with $\tilde{\theta}$ parameters, while $W_{\psi,y}^\mathrm{b}$, and $W_{\psi,u}^\mathrm{b}$ represent the linear parts of the network corresponding to the baseline state reconstruction, and $W_{\psi,y}^\mathrm{a}$, and $W_{\psi,u}^\mathrm{a}$ are utilised for reconstructing the augmented states.

\section{Method} \label{sec:Method}
\subsection{Model-based encoder initialisation}

\subsubsection{LTI state reconstruction}
The matrices $W_{\psi,y}^\mathrm{b}$ and $W_{\psi,u}^\mathrm{b}$ can be initialised using model-based knowledge by utilising the concept of reconstructability. First, let us assume that the available baseline model \eqref{eq:baseline_model} is in \emph{linear time-invariant} (LTI) state-space form as
\begin{subequations}\label{eqs:LTI-SS}
\begin{align}
    x_{\mathrm{b}, k+1} &= A x_{\mathrm{b}, k} + B u_k,\label{eq:x_k+1}\\
    y_k &= Cx_{\mathrm{b}, k+1} + D u_k,\label{eq:y_k}
\end{align}
\end{subequations}
where $y_k\in\mathbb{R}^{n_y}$ now represents the model output corresponding to the baseline component, with the assumption that no measurement noise is present (no contribution of $e_k$).
The effects of measurement and process noise are considered in the next subsection.

From \eqref{eq:x_k+1} and \eqref{eq:y_k}, $\hat{y}_{k+1}$, \dots, $\hat{y}_{k+n}$ can be recursively expressed using only $x_{\mathrm{b},k}$, the input values $u_k$, \dots, $u_{k+n}$ and the system matrices:
\begin{multline}\label{eq:y_k_k+n}
    \underbrace{\begin{bmatrix}
        y_{k+n}\\ y_{k+n-1}\\ \vdots\\ y_{k+1}\\ y_k
    \end{bmatrix}}_{y_k^{k+n}}=\underbrace{\begin{bmatrix}
        CA^n\\ CA^{n-1}\\ \vdots\\ CA\\ C
    \end{bmatrix}}_{\mathcal{O}_n} x_{\mathrm{b},k}+\\ \underbrace{\begin{bmatrix}
        D & CB & CAB & \cdots & CA^{n-1}B\\
        0 & D& CB & \cdots & CA^{n-2}B\\
        \vdots & \vdots &\vdots & \ddots &\vdots\\
        0 & 0 & 0 & \cdots & CB\\
        0 & 0 & 0 & 0 & D\\
    \end{bmatrix}}_{\mathcal{T}_n} \underbrace{\begin{bmatrix}         u_{k+n}\\ u_{k+n-1}\\ u_{k+n-2}\\ \vdots\\ u_k
    \end{bmatrix}}_{u_k^{k+n}}.
\end{multline}

Since the system is time-invariant, shifting back the indices $n$-times can be used for expressing $y_{k-n}^k$, hence
\begin{equation}
    y_{k-n}^k=\mathcal{O}_n x_{\mathrm{b},k-n} + \mathcal{T}_n u_{k-n}^k.\label{eq:y_{k-n}^k}
\end{equation}

From \eqref{eq:y_{k-n}^k}, the state vector can be expressed as
\begin{equation}
    x_{\mathrm{b},k-n}=\mathcal{O}_n^{-1} \left(y_{k-n}^k - \mathcal{T}_n u_{k-n}^k\right),\label{eq:x_k-n}
\end{equation}
where we assume that $\mathcal{O}_n^{-1}$ exists, but since $\mathcal{O}_n\in\mathbb{R}^{(n+1)n_y \times n_x}$, this is not the standard matrix inverse. When the SS representation is observable, $\mathrm{rank}\,\mathcal{O}_n=n_x$ with $ n\geq n_x$, so in that case $\mathcal{O}_n$ always has a left inverse. One approach to compute it is to take the Moore-Penrose pseudoinverse of $\mathcal{O}_n$.

Similarly as in \eqref{eq:y_k_k+n}, $x_{\mathrm{b},k+n}$ can be expressed as
\begin{equation}
    x_{\mathrm{b},k+n} = A^n x_{\mathrm{b},k} + \underbrace{\begin{bmatrix}
        0 & B & AB & \cdots & A^{n-1}B,
    \end{bmatrix}}_{r_n} u_{k}^{k+n},
\end{equation}
then shifting back time, $x_{\mathrm{b},k}$ can be given:
\begin{equation}
    x_{\mathrm{b},k}=A^n  x_{\mathrm{b},k-n} + r_n u_{k-n}^k.\label{eq:xk_subst}
\end{equation}

Using \eqref{eq:xk_subst}, $x_{\mathrm{b},k}$ can be expressed in terms of lagged instances of $u$ and $y$ only, if $x_{\mathrm{b},k-n}$ is substituted by \eqref{eq:x_k-n}:
\begin{equation}
    x_{\mathrm{b},k}=A^n \mathcal{O}_n^{-1} \left(y_{k-n}^k - \mathcal{T}_n u_{k-n}^k\right) + r_n u_{k-n}^k.\label{eq:final_noiseless}
\end{equation}
Note that by rearranging \eqref{eq:final_noiseless}, a direct derivation of $W^\mathrm{b}_{\psi,u}$ and $W^\mathrm{b}_{\psi,y}$ can be given as
\begin{align}
    W^\mathrm{b}_{\psi,u} &= -A^n \mathcal{O}_{n}^{-1} \mathcal{T}_n + r_n,\\
    W^\mathrm{b}_{\psi,y} &= A^n\mathcal{O}_n^{-1}.
\end{align}

If white Gaussian measurement noise is present in the data-generating system, then \eqref{eq:final_noiseless} is a Maximum Likelihood estimator of $x_{\mathrm{b},k}$. However, the presence of process noise in the system dynamics requires a more careful approach to state reconstruction. The next subsection extends the proposed model-based encoder initialisation method to account for the noise structure, incorporating available knowledge about the noise process.

\subsubsection{LTI state reconstruction with noise model}\label{sec:LTI_reconst_noise}
If a noise model is known for the system at hand, then by incorporating it into the LTI baseline model in \eqref{eq:x_k+1} and \eqref{eq:y_k}, the following innovation form of the overall dynamics can be given:
\begin{subequations}
\begin{align}
    x_{\mathrm{b},k+1}&=A x_{\mathrm{b},k} + B u_k + K e_k,\label{eq:x_{k+1}(noise)}\\
    y_k &= C x_{\mathrm{b},k} + D u_k + e_k,\label{eq:y_k(noise)}
\end{align}
\end{subequations}
where $e_k\in\mathbb{R}^{n_y}$ is an i.i.d. zero-mean white noise process with finite variance $\Sigma_e$ and $K\in\mathbb{R}^{n_x \times n_y}$ is a "parameter" matrix which determines how much the process noise affects the state transition.

From \eqref{eq:y_k(noise)}, the noise realisation at time step $k$ can be expressed as
\begin{equation}
    e_k=y_k-C x_{\mathrm{b},k} - D u_k.\label{eq:e_k}
\end{equation}
Then, the state transition can be expressed by substituting \eqref{eq:e_k} into \eqref{eq:x_{k+1}(noise)} as
\begin{equation}
    x_{\mathrm{b},k+1}=\underbrace{\left(A-KC\right)}_{\tilde{A}} x_{\mathrm{b},k} + \underbrace{\left(B-KD\right)}_{\tilde{B}} u_k + K y_k.
\end{equation}

Similarly to \eqref{eq:y_k_k+n}, $y_k^{k+n}$ can be expressed as:
\begin{multline}
        y_k^{k+n} = \tilde{\mathcal{O}}_n x_{\mathrm{b},k} + \tilde{\mathcal{T}}_n u_k^{k+n} +\\ \underbrace{\begin{bmatrix}
        0 & CK & C\tilde{A}K & \cdots & C\tilde{A}^{n-1}K\\
        0 & 0 & CK & \cdots & C\tilde{A}^{n-2}K\\
        \vdots & \vdots &\vdots & \ddots &\vdots\\
        0 & 0 & 0 & \cdots & CK\\
        0 & 0 & 0 & 0 & 0\\
    \end{bmatrix}}_{\tilde{\Lambda}_n} y_k^{k+n}+e_{k}^{k+n},\label{eq:y_kn(noise)}
\end{multline}
where $\tilde{\mathcal{O}}_n$ and $\tilde{\mathcal{T}}_n$ are constructed as $\mathcal{O}_n$ and $\mathcal{T}_n$ in \eqref{eq:y_k_k+n}, but using $\tilde{A}$ instead of $A$ and $\tilde{B}$ instead of $B$, respectively. It seems to be a contradiction that $y_k^{k+n}$ appears both on the left and right side of the equation in \eqref{eq:y_kn(noise)}, but it can be seen that for calculating $y_{k+n}$, output values are needed only up until $y_{k+n-1}$.

As previously mentioned, the time invariance of the system can be utilised as
\begin{equation}
    y_{k-n}^k=\tilde{\mathcal{O}}_n x_{\mathrm{b},k-n} + \tilde{\mathcal{T}}_n u_{k-n}^k + \tilde{\Lambda}_n y_{k-n}^k+e_{k-n}^k.\label{eq:y_k-n-k(noise)}
\end{equation}
Then from \eqref{eq:y_k-n-k(noise)}, the state vector can be expressed as
\begin{equation}
    x_{\mathrm{b},k-n} = \tilde{\mathcal{O}}^{-1}_n \left[ \left(I - \tilde{\Lambda}_n \right) y_{k-n}^k - \tilde{\mathcal{T}}_n u_{k-n}^k - e_{k-n}^k\right],
\end{equation}
where, similarly to the noiseless scenario, $\tilde{O}_n$ always has a left inverse if the $(\tilde{A},\,\tilde{B},\,C,\,D)$ LTI state-space representation is observable and $n$ is chosen sufficiently large.

Similarly to the noiseless scenario, $x_{\mathrm{b},k+n}$ can be expressed in the same way, with the addition of the noise structure as
\begin{multline}
    x_{\mathrm{b},k+n}=\tilde{A}^n x_{\mathrm{b},k} + \underbrace{\begin{bmatrix}
        0 & \tilde{B} & \tilde{A}\tilde{B} & \cdots & \tilde{A}^{n-1}\tilde{B}
    \end{bmatrix}}_{\tilde{r}_n} u_{k}^{k+n} + \\\underbrace{\begin{bmatrix}
        0 & K & \tilde{A}K & \cdots & \tilde{A}^{n-1}K
    \end{bmatrix}}_{\tilde{\lambda}_n} y_k^{k+n},
\end{multline}
then shifting back in time:
\begin{equation}
    x_{\mathrm{b},k}=\tilde{A}^n x_{\mathrm{b},k-n} + \tilde{r}_n u_{k-n}^k + \tilde{\lambda}_n y_{k-n}^k.
\end{equation}

Now we can take the same substitution as in the noiseless case:
\begin{multline}\label{noisy:rec}
    x_{\mathrm{b},k}=\tilde{A}^n \tilde{\mathcal{O}}^{-1}_n \left[ \left(I - \tilde{\Lambda}_n \right) y_{k-n}^k - \tilde{\mathcal{T}}_n u_{k-n}^k- e_{k-n}^k\right] +\\ \tilde{r}_n u_{k-n}^k + \tilde{\lambda}_n y_{k-n}^k.
\end{multline}

However, compared to the noiseless scenario, we can not directly use \eqref{noisy:rec} as the sample path of the noise $\{e_\tau\}_{\tau=k-n}^{k}$ is unknown. However, we can take the conditional expectation of $x_{\mathrm{b},k}$ w.r.t.~the known samples of $u_{k-n}^k$ and $y_{k-n}^k$, which, due to the assumption that $\mathbb{E}\{e_k\}=0$, gives
\begin{multline}\label{noisy:final}
    \mathbb{E}\{x_{\mathrm{b},k}\mid y_{k-n}^k,u_{k-n}^k\}=\\\tilde{A}^n \tilde{\mathcal{O}}^{-1}_n \left[ \left(I - \tilde{\Lambda}_n \right) y_{k-n}^k - \tilde{\mathcal{T}}_n u_{k-n}^k\right] +\\ \tilde{r}_n u_{k-n}^k + \tilde{\lambda}_n y_{k-n}^k,
\end{multline}
which is similar to the noiseless case \eqref{eq:final_noiseless} except the corrector term $\tilde{\lambda}_n$ at the end of the equation. Furthermore, if the generating noise has a Gaussian distribution, i.e., $e_k\sim \mathcal{N}(0,\Sigma_\mathrm{e})$, then \eqref{noisy:final} is a  Maximum Likelihood estimator of $x_{\mathrm{b},k}$.

Then finally, the initialisation of $W^\mathrm{b}_{\psi,u}$, and $W^\mathrm{b}_{\psi,y}$ can be given as
\begin{align}
    W^\mathrm{b}_{\psi,u} &= -\tilde{A}^n\tilde{\mathcal{O}}_n^{-1}\tilde{\mathcal{T}} + \tilde{r}_n,\\
    W^\mathrm{b}_{\psi,y} &= \tilde{A}^n\tilde{\mathcal{O}}_n^{-1} \left(I - \tilde{\Lambda}_n\right) + \tilde{\lambda}_n.
\end{align}

\subsubsection{Generalisation to NL baseline models}
The previously presented initialisation methods considered LTI-SS baseline models. In the general case, when the baseline model is not LTI, an approximated model can be provided based on the local linearisation of the NL model.

Consider an equilibrium point $(x^\ast_\mathrm{b},u^*)\in \mathbb{R}^{n_x} \times \mathbb{R}^{n_u}$, corresponding to
\begin{equation}
    f_\text{base}(\theta_\text{base}, x^\ast_\mathrm{b}, u^\ast) = x_\mathrm{b}^\ast,\quad
    h_\mathrm{base}(\theta_\text{base}, x_\mathrm{b}^\ast, u^\ast) = y^\ast.
\end{equation}

Then, local linearisation around that equilibrium point can be provided by calculating the following Jacobian matrices
\begin{subequations}
\begin{align}
    A^* &= \pdv{f_\mathrm{base}(\theta_\mathrm{b},x_\mathrm{b}^\ast,u^\ast)}{x_\mathrm{b}},\,
    B^* = \pdv{f_\mathrm{base}(\theta_\mathrm{b},x_\mathrm{b}^\ast,u^\ast)}{u},\\
    C^* &= \pdv{h_\mathrm{base}(\theta_\mathrm{b},x_\mathrm{b}^\ast,u^\ast)}{x_\mathrm{b}},\,
    D^* = \pdv{f_\mathrm{base}(\theta_\mathrm{b},x_\mathrm{b}^\ast,u^\ast)}{u}.
\end{align}
\end{subequations}

To get an LTI-SS model, the following variables are introduced: $\tilde{x}_\mathrm{b} = x_\mathrm{b} - x_\mathrm{b}^*$, $\tilde{u} = u - u^*$, and $\tilde{y} = y - y^*$. Then
\begin{subequations}
\begin{align}
    \tilde{x}_{\mathrm{b},k+1} &= A^*\tilde{x}_{\mathrm{b},k} + B^*\tilde{u}_k,\\
    \tilde{y}_k &= C^*\tilde{x}_{\mathrm{b},k} + D^*\tilde{u}_k.
\end{align}
\end{subequations}

After obtaining the system matrices of the linearised system, $W^\mathrm{b}_{\psi,u}$ and $W^\mathrm{b}_{\psi,y}$ can be calculated to reconstruct the initial state $\tilde{x}_{\mathrm{b},i}$ based on $\tilde{u}_{i-n}^i$ and $\tilde{y}_{i-n}^i$. And then $x_{\mathrm{b},i}$ can be calculated\footnote{Alternatively, the bias terms in \eqref{eq:encoder_ann_struct} may be initialised with $x_\mathrm{b}^*$. Utilising these bias terms also enables linearisation around non-equilibrium points, though this requires minor adjustments to the reconstructability map \eqref{noisy:final} to account for the resulting affine terms.} as $x_{\mathrm{b},i}=\tilde{x}_{\mathrm{b},i}+x_\mathrm{b}^*$. %
Keep in mind that the model-based state reconstruction will only be accurate around the equilibrium point $(x_\mathrm{b}^*,u^*)$, but the proposed method is only applied for initialising the parameters of the encoder network. Then, the optimisation can adapt the parameters so that the encoder network can accurately reconstruct the model states, even considering the nonlinear effects.

\subsection{Data-based encoder initialisation}
Alternatively, we propose a data-based method which initialises the encoder as a pre-trained black-box approximation fitted on data available through the baseline model. 
We first generate an approximated dataset by forward simulating the baseline model on $\mathcal{D}_N$, generating $\hat{\mathcal{D}}_N = \{ \left(\hat{y}_k, \hat{x}_{\mathrm{b},k}, u_k\right)\}_{k=1}^N$. If the baseline model is in LTI-SS form, it is sufficient to fit the initial values of $W_{\psi,y}^\mathrm{b}$ and $W_{\psi,u}^\mathrm{b}$ by solving the following \emph{Linear Least Squares} (LLS) problem:
\begin{equation}
    \min_{W_{\psi,y}^\mathrm{b}, W_{\psi,u}^\mathrm{b}} \vert\vert \Phi \begin{bmatrix}
        {W_{\psi,y}^\mathrm{b}}^\top & {W_{\psi,u}^\mathrm{b}}^\top
    \end{bmatrix}^\top-\hat{X}\vert\vert_2^2,
\end{equation}
where $\Phi$ and $\hat{X}$ matrices are given as
\begin{equation}
    \Phi = \begin{bmatrix}
        \hat{y}_{1} & \dots & \hat{y}_{n} & u_{1} & \dots & u_{n}\\
        \vdots & \ddots & \vdots & \vdots & \ddots & \vdots\\
        \hat{y}_{N-n} & \dots & \hat{y}_{N} & u_{N-n} & \dots & u_{N}
    \end{bmatrix},\ 
    \hat{X} = \begin{bmatrix}
        \hat{x}_{\mathrm{b},n} \\ \vdots\\\hat{x}_{\mathrm{b},N}
    \end{bmatrix}.
\end{equation}
The above formulation offers a data-based alternative for the LTI model-based reconstructability map \eqref{eq:final_noiseless}. However, if the applied baseline model is nonlinear, we propose to fit a randomly initialised ANN encoder $\bar \psi$ on $\hat{\mathcal{D}}_N$ by minimising the following loss function
\begin{equation}
    V_\text{enc}(\theta) = \frac{1}{N} \sum_{k=1}^N \left\|\bar \psi_\theta\left(\hat y_{k-n_a}^{k-1}, u_{k-n_b}^{k-1}\right) -\hat{x}_{\mathrm{b},k}\right\|_2^2.
\end{equation}
This method has the benefit of being able to represent the difficult-to-invert initial state estimation of the nonlinear prior models accurately, as we use the full capability of the flexible ANN estimator. We are, however, susceptible to overfitting on the baseline model behaviour and thus can get poor performance during the estimation of the augmentation component.

\section{Simulation Example} \label{sec:Example}
\subsection{Mass Spring Damper}
\begin{figure}
    \centering
    \includegraphics[width=0.7\linewidth]{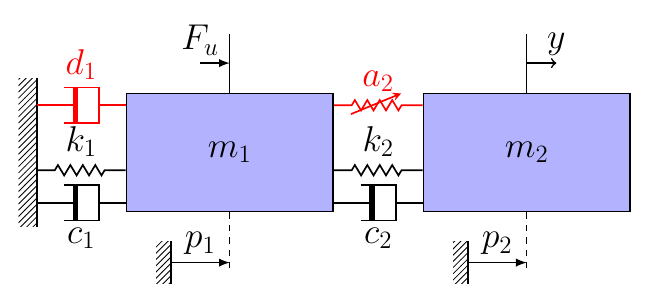}
    \vspace*{-15pt}
    \caption{Nonlinear 2-DOF MSD with cubic damping $d_1$ and cubic spring $a_2$.}
    \label{fig:2dof_msd}
\end{figure}

As a simulation example\footnote{\scriptsize\texttt{https://github.com/JanHHoekstra/Model-Augmentation-Public}}, a nonlinear 2 \emph{degrees-of-freedom} (DOF) \emph{mass-spring-damper} (MSD) is considered, with the nonlinearity being introduced by a hardening spring and damper nonlinearity indicated by the red components in Fig. \ref{fig:2dof_msd}. The parameters of this system are given in Table \ref{tab:params_msd}. The states associated with the system representation are the positions $p_i$ and the velocities $\dot p_i$ of the masses $m_1$ and $m_2$. The hardening nonlinearities are described as cubic terms with parameters $d_1$ and $a_2$. The position $p_2$ is the measured output, and the input force $u$ acts on the first mass $m_1$.

We obtain the data for model estimation by applying \emph{4th~order Runge-Kutta} (RK4) based numerical integration in steps of $T_i = 0.01$~$\mathrm{s}$ on the 2-DOF MSD system with a sampling time of~$T_\mathrm{s}=0.1$~$\mathrm{s}$ (i.e., $10$ integration steps for every sample). The system is simulated for a zero-order hold input signal $F_u$ generated by a DT multisine $u_{k}$ with 1666 frequency components in the range $[0, 5]$ Hz with a uniformly distributed phase in $ [0, 2\pi)$. The sampled output measurements $y_k$ are perturbed by additive noise $e_k$ described by a DT white noise process with $e_k\sim\mathcal{N}(0,\sigma_\mathrm{e}^2)$. Here, $\sigma_\text{e}$ is chosen so that the \emph{signal-to-noise ratio} (SNR) is equal to 20 dB

After removing the transient due to initial conditions, the sampled output $y_k$ for the input~$u_{k}$ is collected in the data set $\mathcal{D}_N$. Separate data sets are generated for estimation, validation, and testing with sizes $N_{\text{est}} = 2\cdot10^4$, $N_{\text{val}} = 10^4$, $N_{\text{test}} = 10^4$, respectively.

\begin{table}[h]
    \centering
    \caption{Physical parameters of the 2-DOF MSD system.}
    \vspace*{-6pt}
    \label{tab:params_msd}
    \begin{tabular}{c||r|r}
        Parameter & Body 1 & Body 2 \\
        \hline
        Mass $m_i$ & $0.5\,\mathrm{kg}$ & $0.4\,\mathrm{kg}$ \\
        Spring constant $k_i$ & $100\,\mathrm{N/m}$ & $100\,\mathrm{N/m}$ \\
        Damping $c_i$ & $0.50\,\mathrm{N\,s/m}$ & $0.50\,\mathrm{N\,s/m}$ \\
        Hardening $a_i$ & - & $1000\,\mathrm{N/m^3}$ \\
        NL damping $d_i$ & $0.1\,\mathrm{N\,s^3/m^3}$ & - \\
    \end{tabular}
\end{table}

\subsection{Baseline Model}
We consider a nonlinear baseline model with the same components as the system with the parameters from Table~\ref{tab:params_msd} but with the nonlinear damping term reduced to  $0.05\,\mathrm{N\,s^3/m^3}$. We evaluate this baseline model by the simulation \emph{root mean square error} (RMSE) over the test data:
\begin{equation}
    \text{RMSE} = \sqrt{\frac{1}{N}\sum_{k=1}^{N}\left|\left| y_k - \yestk \right|\right| ^2 _2}.
\end{equation}
The simulation RMSE of the nonlinear baseline model over the test data is $0.040$ while its linearisation around the $0$ point, used for the model-based initialisation method, has a simulation RMSE of $0.21$.

\subsection{Model augmentation}
We augment the baseline model, with the general augmentation structure \eqref{eq:augmentation_structure} configured as a static parallel augmentation \citep{HOEKSTRA2025}:
\begin{equation}
    \xbasekplus =  f_\text{base}\left(\xbasek, \uk\right) + f_\text{aug}\left(\xbasek, \uk\right)
\end{equation}
The encoder $\psi_\theta$ is structured as a ResNet, as is the augmentation learning component $f_\text{aug}$. This learning component is initialised with zeros in the final layer as described in \citep{HOEKSTRA2025}. Both ResNets are estimated jointly to minimise \eqref{eq:loss_function} with the Adam optimiser \citep{Kingma2015}. The hyperparameters for the ResNets and the Adam optimisation are described in Table~\ref{tab:Hyperparam}. For the data-based encoder initialisation, we use the jax-sysid algorithm \citep{Bem25}. This algorithm first performs a number of Adam epochs to initialise for the following \emph{bound-constrained limited-memory Broyden-Fletcher-Goldfarb-Shanno} (L-BFGS-B) algorithm \citep{liu1989}. The jax-sysid algorithm has been shown to be computationally efficient for the identification of NL-SS systems \citep{Bem25}. The hyperparameters for the jax-sysid algorithm are described in Table~\ref{tab:Hyperparam}.

\subsection{Monte Carlo analysis}\label{sec:nonlinear_encoder_results}
\begin{figure}
    \centering
    \includegraphics[width=1.0\linewidth]{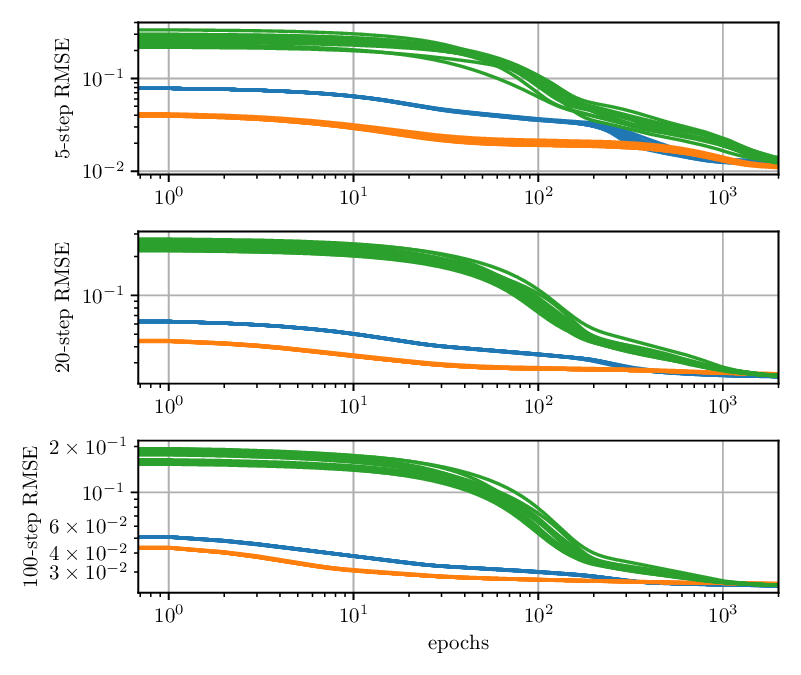}
    \vspace*{-15pt}
    \caption{Validation loss for 5, 10 and 100 step ahead RMSE over the fitting epochs for the three encoder initialisation methods: model-based (blue), data-based (orange) and randomly initialised (green).}
    \label{fig:Loss_function_monte_carlo}
\end{figure}

We perform a Monte Carlo analysis of the loss function convergence by fitting 10 augmented models for the model-based, data-based and random initialisation method each. For all three methods, the parameters that were left random, the first layer of the parallel augmentation and the random encoder initialisation method, are different realisations for each of these 10 models. We evaluate the estimated models on the validation data with the T-step ahead RMSE loss, which is the square root of the loss function \eqref{eq:mse_loss}. We evaluate this metric for T equals 5, 20, and 100 steps ahead prediction over the training epochs as shown in Fig.~\ref{fig:Loss_function_monte_carlo}.

First, the figure shows a clear benefit of the model and data-based encoder initialisation methods over the random initialisation with respect to better starting values. This benefit is more pronounced for shorter T-step ahead predictions. This is expected behaviour, as the decaying impact of the initial state results in the correct identification of the system dynamics through the augmentation becomes important. Second, we also notice that the data-based and to a lesser extent the model-based method, have a faster convergence for the first 1000 epochs, not only for the shorter 5-step ahead prediction measure, but also the 20 and 100-step ahead prediction measure. This indicates that the more accurate initialisation of the encoder results in faster convergence. This could be explained by the joint identification not needing to fight between the encoder and augmentation learning optimisation, but instead being able to learn the augmentation while fine-tuning the encoder. Future research will further explore the reasons for this convergence behaviour.

\begin{figure}
    \centering
    \includegraphics[width=1.0\linewidth]{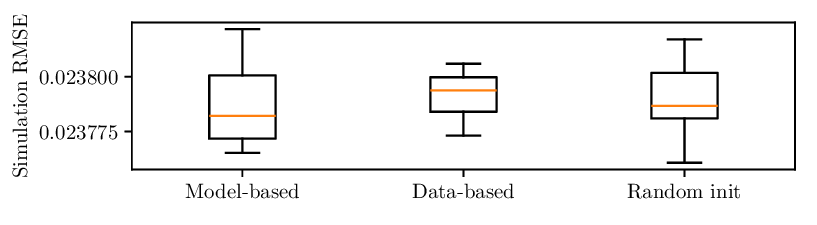}
    \vspace*{-20pt}
    \caption{Distribution of simulation RMSE over the test data for 10 different model for each encoder initialisation method.}
    \label{fig:box_plot}
\end{figure}

The distribution of the final simulation RMSE scores over the test data is shown in Fig.~\ref{fig:box_plot}. Here we can see that for the considered system and baseline model, the final RMSE scores are comparable between the three methods.

Finally, we note the computation time of the encoder initialisation methods. The random initialisation is instantaneous. The model-based method requires an Moore-Penrose inverse and otherwise only matrix multiplications and additions. This has a minimal overhead, in the used implementation it is approximately $1.5$ ms. The data-based method computation time will depend on the optimisation algorithm and the hyperparameters selected for it. The currently used naïve implementation used has a computational time of approximately $4.8$ s. This could likely be significantly reduced by fine tuning the hyperparameters for the optimisation. However, it will remain slower than the model-based method. In applications where computation time is key, e.g., online learning, the model-based initialisation approach provides a compromise between computational efficiency and accuracy. For other scenarios, the data-based initialisation technique might be preferred for nonlinear baseline models.

\begin{table}[t]
    \centering
    \caption{Hyperparameters for identifying the augmented models with Adam algorithm.}
    \vspace*{-8pt}
    \begin{tabular}{c|c|c|c|c|c|c}
    \!hidden layers\!&\!nodes\!&\!$n_u$ $n_y$\!&\!$n_a$ $n_b$\!&\!$T$\!&\!epochs\!&\!batch size\!\\
    \hline
    2 & 16 & 1 & 9 & 200 & 2000 & 3000 \\
    \end{tabular}
    \label{tab:Hyperparam}
\end{table}

\begin{table}[t]
    \centering
    \caption{Hyperparameters for jax-sysid algorithm for data-based initialisation.}
    \vspace*{-8pt}
    \begin{tabular}{c|c|c|c|c|c|c}
    \! Adam epochs \!&\! L-BFGS-B epochs\!&\! memory size\!\\
    \hline
    50 & 200 & 50 \\
    \end{tabular}
    \label{tab:Hyperparam_jax_sysid}
\end{table}

\section{Conclusion} \label{sec:Conclusion}
This paper has proposed methods to initialise the encoder to reconstruct the initial state for simulated identification criteria. We have derived initialisation based on both linear and nonlinear baseline models. The initialisation methods using the nonlinear baseline model have been shown to perform better compared to the random initialisation previously used in the literature in terms of convergence time. This is achieved while maintaining similar final accuracy as compared to the random black-box initialisation currently used in the literature. In future research, we will look into reasoning for the convergence behaviour, nonlinear noise models, and different forms of augmentations.

\bibliography{references}

@article{Schoukens2019,
  author =        {Johan Schoukens and Lennart Ljung},
 journal={IEEE Control Systems Magazine}, 
  pages =         {28-99},
  title =         {Nonlinear System Identification: A User-Oriented Road Map},
  volume =        {39},
  year =          {2019},
}

@article{suykens1995,
  author =        {Suykens, Johan AK and De Moor, Bart LR and
                   Vandewalle, Joos},
  journal =       {International Journal of Control},
  pages =         {129--152},
  title =         {Nonlinear system identification using neural state space models, applicable to robust control design},
  volume =        {62},
  year =          {1995},
}

@inproceedings{Schoukens2021,
  author =        {Schoukens, Maarten},
  booktitle =     {Proc. of the European Control Conference},
  pages =         {1913-1918},
  title =         {Improved Initialization of State-Space Artificial Neural Networks},
  year =          {2021},
}

@article{ljung2010perspectives,
  author =        {Ljung, Lennart},
  journal =       {Annual Reviews in Control},
  pages =         {1--12},
  publisher =     {Elsevier},
  title =         {Perspectives on system identification},
  volume =        {34},
  year =          {2010},
}

@article{sun2020comprehensive,
  author =        {Sun, Bei and others},
  journal =       {Journal of Process Control},
  pages =         {30--43},
  publisher =     {Elsevier},
  title =         {A comprehensive hybrid first principles/machine
                   learning modeling framework for complex industrial
                   processes},
  volume =        {86},
  year =          {2020},
}

@inproceedings{gotte2022composed,
  author =        {G{\"o}tte, Ricarda-Samantha and Timmermann, Julia},
  booktitle =     {Proc. of the 3rd International Conference on
                   Artificial Intelligence, Robotics and Control},
  organization =  {},
  pages =         {67--76},
  title =         {Composed physics-and data-driven system
                   identification for non-autonomous systems in control
                   engineering},
  year =          {2022},
}

@article{Groote2022,
  author =        {Wannes De Groote and others},
  journal =       {IEEE/ASME Transactions on Mechatronics},
  pages =         {103-114},
  publisher =     {Institute of Electrical and Electronics Engineers
                   Inc.},
  title =         {Neural Network Augmented Physics Models for Systems
                   with Partially Unknown Dynamics: Application to
                   Slider-Crank Mechanism},
  volume =        {27},
  year =          {2022}
}

@article{Shah2022,
  author =        {Parth Shah and others},
  journal =       {Chemical Engineering Journal},
  pages =         {135643},
  publisher =     {Elsevier},
  title =         {Deep neural network-based hybrid modeling and
                   experimental validation for an industry-scale
                   fermentation process: Identification of time-varying
                   dependencies among parameters},
  volume =        {441},
  year =          {2022},
}

@article{Retzler2024,
  author =        {András Retzler and others},
  journal =       {Data-Centric Engineering},
  pages =         {e12},
  publisher =     {Cambridge University Press},
  title =         {Learning-based augmentation of physics-based models:
                   an industrial robot use case},
  volume =        {5},
  year =          {2024},
}

@article{Kingma2015,
  author =        {Kingma, Diederik P},
  journal =       {arXiv preprint arXiv:1412.6980},
  title =         {Adam: A method for stochastic optimization},
  year =          {2014},
}

@article{HOEKSTRA2025,
title = {Learning-based model augmentation with LFRs},
journal = {European Journal of Control},
pages = {101304},
year = {2025},
author = {Jan H. Hoekstra and Chris Verhoek and Roland Tóth and Maarten Schoukens},
}

@book{spong2006robot,
  title={Robot Modeling and Control},
  author={Spong, M.W. and Hutchinson, S. and Vidyasagar, M.},
  edition={2nd},
  year={2020},
  publisher={Wiley}
}

@book{isidori1985nonlinear,
  title={Nonlinear Control Systems},
  author={Isidori, A.},
  series={Communications and Control Engineering},
  edition={3rd},
  year={1995},
  publisher={Springer London}
}

@phdthesis{kessels2025ai,
    type = {Phd {Thesis}},
    title = {{AI}-based {Model} {Updating} for {Nonlinear} {Dynamical} {Systems}},
	school = {Eindhoven University of Technology},
    author={Kessels, Bas Mats},
    year={2025}
}

@article{ramkannan2023initialization,
  title={Initialization Approach for Nonlinear State-Space Identification via the Subspace Encoder Approach},
  author={Ramkannan, Rishi and Beintema, Gerben I and T{\'o}th, Roland and Schoukens, Maarten},
  journal={IFAC-PapersOnLine},
  volume={56},
  number={2},
  pages={5146--5151},
  year={2023},
  publisher={Elsevier}
}

@phdthesis{beintema2024data,
    type = {Phd {Thesis}},
    title = {Data–driven {Learning} of {Nonlinear} {Dynamic} {Systems}: {A} {Deep} {Neural} {State}–{Space} {Approach}},
	shorttitle = {Data–driven {Learning} of {Nonlinear} {Dynamic} {Systems}},
	school = {Eindhoven University of Technology},
    author={Beintema, Gerben Izaak},
    year={2024}
}

@article{Bem25,
    author={A. Bemporad},
    title={An {L-BFGS-B} approach for linear and nonlinear system identification under $\ell_1$ and group-Lasso regularization},
    journal = {IEEE Transactions on Automatic Control},
    year={2025},
    volume={70},
    number={7},
    pages={4857-4864}
}

@article{liu1989,
  title = {On the Limited Memory {{BFGS}} Method for Large Scale Optimization},
  author = {Liu, Dong C. and Nocedal, Jorge},
  year = {1989},
  journal = {Mathematical Programming},
  volume = {45},
  pages = {503--528},
}

@article{masti2021learning,
  title={Learning nonlinear state--space models using autoencoders},
  author={Masti, Daniele and Bemporad, Alberto},
  journal={Automatica},
  volume={129},
  pages={109666},
  year={2021},
  publisher={Elsevier}
}

\end{document}